\documentclass[a4paper]{article}
\usepackage{INTERSPEECH2020}
\usepackage{amsmath,graphicx}
\usepackage{booktabs}
\usepackage{tabularx}
\usepackage{bm}

\usepackage{INTERSPEECH2020}

\title{End-to-End Multi-Task Denoising for the Joint Optimization of Perceptual Speech Metric}
\name{Jaeyoung Kim, Mostafa El-Khamy$^1$, Jungwon Lee$^1$}
\address{
  $^1$ Samsung Semiconductor Inc. Multimedia R\&D, USA}
\email{jaeykim@google.com, \{mostafa.e,jungwon2.lee\}@samsung.com}

\begin{document}

\maketitle
\begin{abstract}
Although supervised learning based on a deep neural network has recently achieved substantial improvement on speech enhancement, the existing schemes have either of two critical issues: spectrum or metric mismatches. The spectrum mismatch is a well known issue that any spectrum modification after short-time Fourier transform (STFT), in general, cannot be fully recovered after inverse short-time Fourier transform (ISTFT). The metric mismatch is that a conventional mean square error (MSE) loss function is typically sub-optimal to maximize perceptual speech measure such as signal-to-distortion ratio (SDR), perceptual evaluation of speech quality (PESQ) and short-time objective intelligibility (STOI). This paper presents a new end-to-end denoising framework. First, the network optimization is performed on the time-domain signals after ISTFT to avoid the spectrum mismatch. Second, three loss functions based on SDR, PESQ and STOI are proposed to minimize the metric mismatch. The experimental result showed the proposed denoising scheme significantly improved SDR, PESQ and STOI performance over the existing methods. Moreover, the proposed scheme also provided good generalization performance over generative denoising models on the perceptual speech metrics not used as a loss function during training.
\end{abstract}

\section{Introduction}
\label{sec:intro}

In recent years, deep neural networks have shown great success in speech enhancement compared with traditional statistical approaches.
%~\cite{ephraim1984speech,ephraim1985speech, cohen2001speech}. 
Neural networks directly learn complicated nonlinear mapping from noisy speech to clean one only by referencing data without any prior assumption.

Spectral mask estimation is a popular supervised denoising method that predicts a time-frequency mask to obtain an estimate of clean speech by scaling the noisy spectrum. There are numerous types of spectral mask estimation techniques depending on how to define mask labels. For example, authors in~\cite{narayanan2013ideal} proposed the ideal binary mask (IBM) as a training label, where it is set to be zero or one depending on the signal to noise ratio (SNR) of the noisy spectrum. The ideal ratio mask (IRM)~\cite{wang2014training} and the ideal amplitude mask (IAM)~\cite{erdogan2015phase} provided non-binary soft mask labels to overcome the coarse label mapping of IBM. The phase sensitive mask (PSM)~\cite{erdogan2015phase} considers the phase spectrum difference between clean and noisy signals, in order to correctly maximize the signal to noise ratio (SNR).  

Generative models, such as generative adversarial networks (GANs) suggested an alternative to supervised learning. In speech enhancement GAN (SEGAN)~\cite{pascual2017segan}, a generator network is trained to output a time-domain denoised signal that can fool a discriminator from a true clean signal. TF-SEGAN  \cite{soni2018time} extended SEGAN to use a time-frequency mask. 

However, all the schemes described above suffer from at least one of two critical issues: metric  mismatch or spectrum mismatch. Signal-to-distortion ratio (SDR), perceptual evaluation of speech quality (PESQ) and short-time objective intelligibility (STOI) are the most well-known perceptual speech metrics. %ME: spell out SDR, PESQ
 The typical mean square error (MSE) criterion popularly used for spectral mask estimation is not optimal to maximize them. For example, decreasing the mean square error of noisy speech signals often degrades SDR, PESQ or STOI due to different weighting of the frequency components or non-linear transforms involved in those metrics. The spectrum mismatch is a well known issue: any modification of the spectrum after the short-time Fourier transform (STFT), in general, cannot be fully recovered after inverse short-time Fourier transform (ISTFT)~\cite{allen1977unified}.
Therefore, spectral mask estimation and other alternatives optimized in the spectrum domain always have a potential risk of performance loss.

This paper presents a new end-to-end multi-task denoising scheme with the following contributions. First, the proposed framework presents three loss functions:
\begin{itemize}
	\item SDR loss function: The SDR metric is used as a loss function. The scale-invariant term in the SDR metric is incorporated as a part of training, which provided a significant SDR boost.  
	\item PESQ loss function: The PESQ metric is redesigned to be usable as a loss function. 
	\item STOI loss function: The proposed STOI loss function modified the original STOI metric by allowing 16 kHz sampled acoustic signal to be used without resampling.    
\end{itemize}
The proposed multi-task denoising scheme combines loss functions mentioned above for the joint optimization of SDR, PESQ and STOI metrics.  
%ME: in many instances you mention SDR loss, PESQ loss, I corrected some of this above, but I think it should be corrected over all the rest of the document to be "`the SDR los', "the PESQ loss"
Second, a denoising network still predicts a time-frequency mask, but the network optimization is performed after ISTFT in order to avoid spectrum mismatch. The evaluation result showed that the proposed framework provided large improvement on SDR, PESQ and STOI. Moreover, the proposed scheme also showed good generalization on the unseen metrics as in Table~\ref{tab:ge}.
\section{The Proposed Framework}
\label{sec:end_to_end}
\begin{figure}[ht]
%\begin{figure*}
%\vskip -0.2in
\begin{center}
\centerline{\includegraphics[width=55mm]{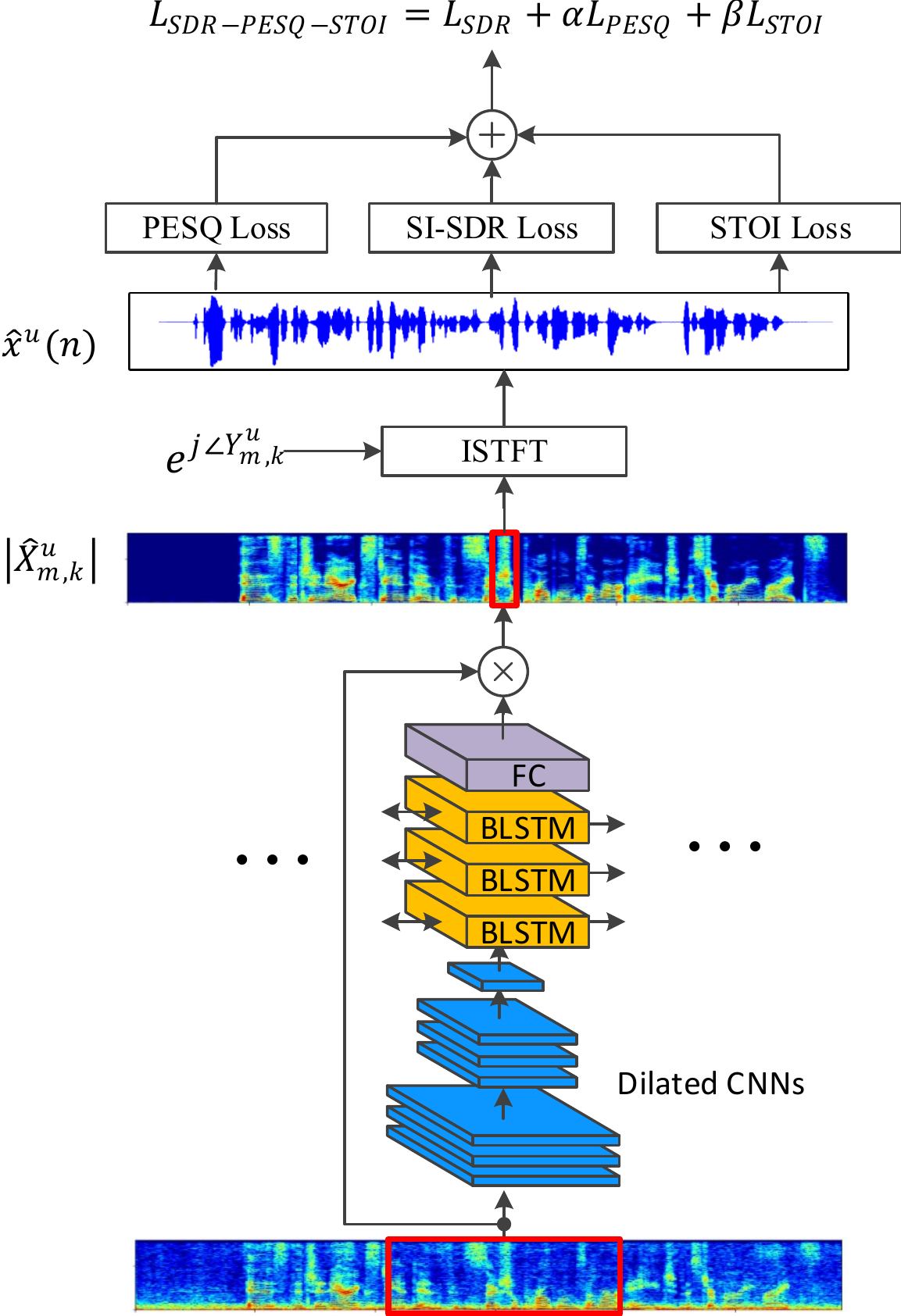}}
%\centerline{\includegraphics[width=\textwidth]{figs/denoising_framework-cropped.pdf}}
\caption{Illustration of End-to-End Multi-Task Denoising based on CNN-BLSTM}
\label{fig:ete}
\end{center}
\vskip -0.4in
\end{figure} 
Figure~\ref{fig:ete} describes the proposed end-to-end multi-task denoising framework. The underlying model architecture is composed of convolutional layers and bi-directional LSTM. The spectrogram formed by 11 frames of the noisy amplitude spectrum $|Y^u_{m,k}|$ is the input to the convolutional layers. $u$ is an utterance index, $m$ is a frame index and $k$ is a frequency index. The dilated convolution with rate of 2 and 4 is applied to the second and third layers, respectively, in order to increase kernel's coverage on frequency bins. Dilation is only applied to the frequency dimension because time correlation will be learned by bi-drectional LSTMs. Griffin-Lim ISTFT~\cite{griffin1984signal} is applied to the synthesized complex spectrum $\hat{X}^u_{m,k}$ to obtain time-domain denoised output $\hat{x}^u(n)$. Three proposed loss functions are evaluated based on $\hat{x}^u(n)$ and therefore, they are free from the spectrum mismatch.  

\subsection{SDR Loss Function}
\label{sec:si-sdr}
\begin{figure}[ht]
%\begin{figure*}
\vskip -0.1in
\begin{center}
\centerline{\includegraphics[width=28mm]{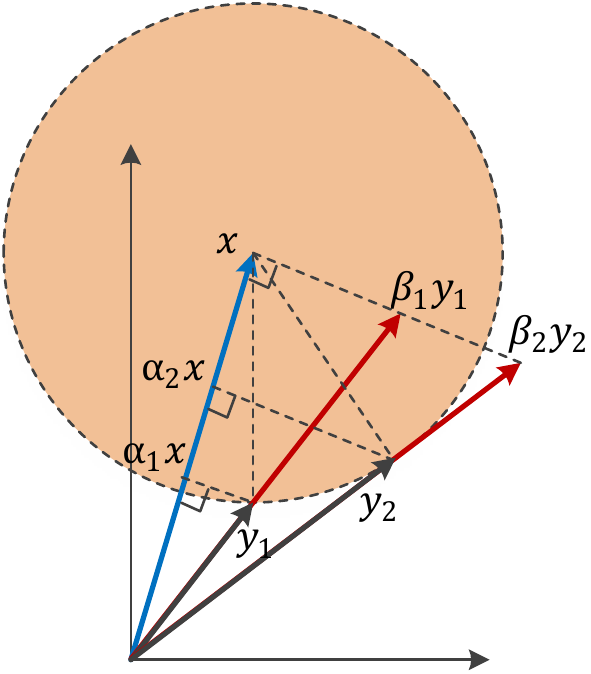}}
%\centerline{\includegraphics[width=\textwidth]{figs/denoising_framework-cropped.pdf}}
\caption{SDR comparison between two noisy signals with the same SNR}
\label{fig:si_sdr}
\end{center}
\vskip -0.2in
\end{figure}

Unlike SNR, the definition of SDR is not unique. There are at least two popularly used SDR definitions. The most well known Vincent's definition~\cite{vincent2006performance} is given by
\begin{equation}
\textrm{SDR}^u=10\log_{10}\frac{|| x^u_{target} ||^2}{|| e^u_{noise} + e^u_{artif} ||^2}
\label{eq:sdr}
\end{equation}
$x^u_{target}$ and $e^u_{noise}$ can be found by projecting the denoised signal $\hat{x}^u(n)$ into the clean and noise signal domains, respectively. $e^u_{artif}$ is a residual term. They can be formulated as follows:
\begin{eqnarray}
  x^u_{target}&=&\frac{(x^u)^T\hat{x}^u}{||x^u||^2}x^u \label{eq:t} \\
  e^u_{noise} &=&\frac{(n^u)^T\hat{x}^u}{||n^u||^2}n^u \label{eq:n}\\ 
  e^u_{artif} &=&\hat{x}^u-\left( \frac{(x^u)^T\hat{x}^u}{||x^u||^2}x^u + \frac{(n^u)^T\hat{x}^u}{||n^u||^2}n^u \right) \label{eq:a}
\end{eqnarray}
Substituting Eq.(\ref{eq:t}), (\ref{eq:n}) and (\ref{eq:a}) into Eq.(\ref{eq:sdr}), the rearranged SDR is given by
\begin{eqnarray}
\label{eq:sdr2}
\textrm{SDR}^u
     % &=& 10\log_{10}\frac{|| \frac{(x^u)^T\hat{x}^u}{||x^u||^2}x^u ||^2}{||\frac{(x^u)^T\hat{x}^u}{||x^u||^2}x^u-\hat{x}^u ||^2} \nonumber \\
      &=& 10\log_{10}\frac{|| \alpha^u x^u ||^2}{|| \alpha^u x^u-\hat{x}^u ||^2}
 \end{eqnarray}
where $\alpha^u = \operatorname*{arg\,min}_{\alpha} ||\alpha x^u - \hat{x}^u || =\frac{(x^u)^T\hat{x}^u}{||x^u||^2}x^u$. Eq. (\ref{eq:sdr2}) coincides with scale-invariant SDR (SI-SDR), which is another popularly used SDR definition~\cite{roux2018sdr}. In the general multiple source denoising problems, SDR and SI-SDR do not match each other. However, for the single source denoising problem, we can use them interchangeably.
SDR loss function is defined as mini-batch average of Eq.(\ref{eq:sdr2}):
\begin{equation}
\label{eq:si_sdr}
\textrm{L}_{\textrm{SDR}}=\frac{1}{B}\sum_{u=0}^{B-1} 10\log_{10}\frac{|| \alpha^u x^u ||^2}{|| \alpha^u x^u-\hat{x^u} ||^2}
\end{equation}
where $\alpha$ is $\frac{(x^u)^T\hat{x^u}}{||x^u||^2}$.
%Compared with the conventional MSE loss function, scale-invariant $\alpha$ is included as a training variable in Eq.(\ref{eq:si_sdr}). 
Figure~\ref{fig:si_sdr} illustrates why training $\alpha$ is important to maximize the SDR metric. For two noisy signals $y_1$ and $y_2$, they have the same SNR of $\frac{||x||^2}{||x - y_1||^2} = \frac{||x||^2}{||x - y_2||^2}$ because they are positioned at the same circle centered at the clean signal $x$. However, their SDRs are different. For $y_1$, SDR is $\frac{||\alpha_1 x ||^2}{||\alpha_1 x - y_1||^2}$, which is $\frac{||x ||^2}{||x - \beta_1 y_1||^2}$ by geometry. By the same way, SDR for $y_2$ is  $\frac{||x ||^2}{||x - \beta_2 y_2||^2}$. Clearly, $y_1$ is better than $y_2$ in terms of SDR metric, but the MSE criterion cannot distinguish between them because they have the same SNR. 

\subsection{PESQ Loss Function}
\label{sec:pesq}
\begin{figure}[ht]
%\begin{figure*}
\vskip -0.2in
\begin{center}
\centerline{\includegraphics[width=70mm]{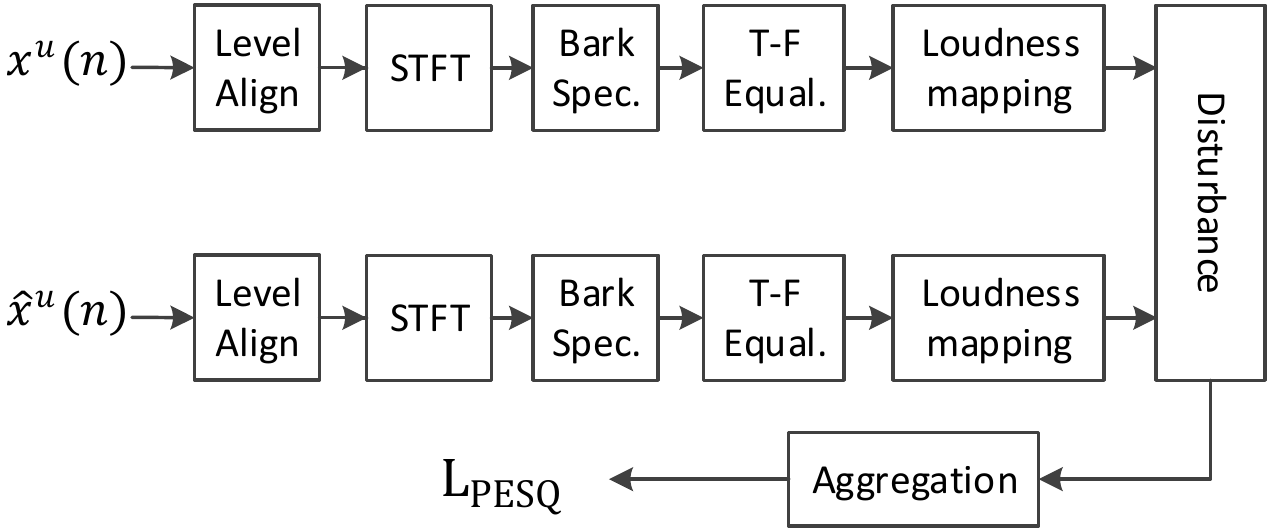}}
%\centerline{\includegraphics[width=\textwidth]{figs/denoising_framework-cropped.pdf}}
\caption{Block Diagram for the PESQ loss function: $x^u(n)$ and $\hat{x}^u(n)$ are clean and denoised time-domain signals.}
\label{fig:pesq}
\end{center}
\vskip -0.4in
\end{figure}

PESQ~\cite{rix2001perceptual} is defined by ITU-T recommendation P.862 for the objective speech quality evaluation. 
%Its value ranges from -0.5 to 4.5 and the higher PESQ means better perceptual quality. 
%Although SDR and PESQ have high correlation, it is frequently observed that signals with smaller SDR have higher PESQ than the ones with the higher SDR. 
%For example, an acoustic signal with high reverberation would have low SNR or SDR, but PESQ can be much better because time-frequency equalization in PESQ can compensate the most of channel fading effects. It is just one example and there are a lot of operations that make PESQ behave much differently from SDR. Therefore, SDR loss function cannot effectively optimize PESQ. 
%In this section, a new PESQ loss function is designed based on the PESQ metric. 
Figure~\ref{fig:pesq} shows the block diagram to find PESQ loss function $\textrm{L}_{\textrm{PESQ}}$. The overall procedure is similar to the PESQ system in ~\cite{rix2001perceptual}. There are three major modifications. First, the IIR filter in the PESQ calculation is removed. The reason is that time evolution of IIR filter is too deep to apply back-propagation. Second, delay adjustment routine is not necessary because training data pairs were already time-aligned. Third, bad-interval iteration is removed. As long as clean and noisy data pairs are time-aligned, there's no substantial impact on PESQ from removing it. Each block in Figure~\ref{fig:pesq} is briefly explained below:

\textbf{\underline{Level Alignment}}: The average power of clean and noisy speech ranging from 300Hz to 3KHz are set to be predefined power value, $10^7$. 

\textbf{\underline{Bark Spectrum}}: The Bark spectrum block is to convert linear scale frequency bins into the Bark scale which is roughly equivalent to a logarithmic scale. 
The mapped bark spectrum power can be formulated as follows:
\begin{equation}
B^u_{c,m,i}=\frac{1}{I_{i+1}-I_i}\sum_{k=I_i}^{I_{i+1}-1}|X^u_{m,k}|^2
\end{equation}
where $I_i$ is the start of linear frequency bin number for the $i^{th}$ bark spectrun, $B^u_{c,m,i}$ is $i^{th}$ bark spectrum power of clean speech. All the positive linear frequency bins were mapped to 49 Bark spectrum bins. $B^u_{n,m,i}$ is a bark spectrum power of noisy speech and can be also found in the similar manner.

\textbf{\underline{Time-Frequency Equalization}}: Each Bark spectrum bin of clean speech is compensated by the average power ratio between clean and noisy Bark spectrum as follows
\begin{equation}
E^u_{c,m,i}=\frac{P^u_{n,i}+c_1}{P^u_{c,i}+c_1}B^u_{c,m,i}
\end{equation}
where $P^u_{n,i}=\frac{1}{M_u}\sum_{m=0}^{M_u-1} B^u_{n,m,i}S^u_{n,m,i}$ and $P^u_{c,i}=\frac{1}{M_u}\sum_{m=0}^{M_u-1} B^u_{c,m,i}S^u_{c,m,i}$. $S^u_{c,m,i}$ and $S^u_{n,m,i}$ are silence masks for clean and noisy speech, respectively. They are fixed arrays of thresholds intended to avoid frequency compensation for the region where clean or noisy speech powers were small. 49 entries of silence masks are defined in PESQ source code~\cite{itut802}. $c_1$ is a small constant whose main purpose is to avoid division by zero. It is currently set as 1000. 
A noise bark spectrum is also similarly compensated for each frame.
%After frequency equalization, the short-term gain variation of a noisy bark spectrum is also compensated for each frame:
%\begin{eqnarray}
%S^u_m &=& \frac{G^u_{c,m}+c_2}{G^u_{n,m}+c_2} \label{eq:tf1}\\
%R^u_m &=& 0.2S^u_{m-1}+0.8S^u_{m}\label{eq:tf2} \\
%E^u_{n,m,i} &=& R^u_{m}B^u_{n,m,i} \label{eq:tf3}
%\end{eqnarray}
%where $G^u_{n,m}=\sum_{i=0}^{I-1}B^u_{n,m,i}$, $G^u_{c,m}=\sum_{i=0}^{I-1}E^u_{c,m,i}$ , $I$ is the size of Bark spectrum bins, and $c_2$ is a constant. 

\textbf{\underline{Loudness Mapping}}: The power densities are transformed to a Soner loudness scale using Zwicker's law~\cite{zwicker1967ohr}:
\begin{equation}
L^u_{x,m,i}=S_i\left(\frac{P_{0,i}}{0.5}\right) \left[ \left(0.5 + 0.5\frac{E^u_{x,m,i}}{P_{0,i}}\right) -1\right]
\end{equation}
where $P_{0,i}$ is the absolute hearing threshold, $S_i$ is the loudness scaling factor, and r is Zwicker power and x can be c (clean) or n (noisy).

\textbf{\underline{Disturbance Processing}}: The raw disturbance is difference between clean and noisy loudness densities with following operations:
\begin{eqnarray}
D^u_{m,i}=\max \left( L^u_{c,m,i} - L^u_{n,m,i} - \textrm{DZ}^u_{m,i},0 \right)&& \nonumber \\
          + \min \left( L^u_{c,m,i} - L^u_{n,m,i} + \textrm{DZ}^u_{m,i},0 \right)&&
\end{eqnarray} 
where $\textrm{DZ}^u_{m,i} = 0.25\min(L^u_{c,m,i},L^u_{n,m,i})$.
If absolute difference between clean and noisy loudness densities are less than 0.25 of minimum of two densities, raw disturbance becomes zero. From raw disturbance, symmetric frame disturbance is given by
\begin{equation}
\textrm{FD}^u_{m}=\sqrt{\frac{1}{\sum_{i=0}^{I-1}w_i}\sum_{i=0}^{I-1}\left( w_i D^u_{m,i} \right)^2}
\end{equation}
where $w_i$ is predefined weighting for bark spectrum bins. It is also defined in PESQ source code~\cite{itut802}. Asymmetric frame disturbance has additional scaling and thresholding steps as follows:
\begin{eqnarray}
h^u_{m,i}&=& \left( \frac{B^u_{n,m,i}+50}{B^u_{c,m,i}+50}   \right)^{1.2} \\
h^u_{m,i}&=& \bigg\{
\begin{array}{l}
12,  \textrm{   if   } h^u_{m,i} > 12  \\
0,   \textrm{   if } h^u_{m,i} < 3
\end{array}
\end{eqnarray} 
\begin{equation}
\textrm{AFD}^u_{m} =\sqrt{\frac{1}{\sum_{i=0}^{I-1}w_i}\sum_{i=0}^{I-1}\left( w_i D^u_{m,i} h^u_{m,i}\right)^2}
\end{equation}
\textbf{\underline{Aggregation}}: PESQ loss function $L_{\textrm{PESQ}}$ can be found as follows:
\begin{equation}
L_{\textrm{PESQ}}=4.5-0.1d_{sym}-0.0309d_{asym}
\end{equation}
$d_{sym}$ can be found from two stage averaging:
\begin{eqnarray}
\textrm{PSQM}_s^u=\sqrt[6]{\frac{1}{20}\sum_{i=0}^{19}\left( \textrm{FD}^u_{10s+i} \right)^6} \\
d_{sym}=\frac{1}{B}\sum_{u=0}^{B-1} \sqrt{\frac{1}{S^u}\sum_{s=0}^{S^u-1}\left( \textrm{PSQM}_s^u \right)^2}
\end{eqnarray}
where $S^u$ is $\lfloor \frac{M^u}{10} \rfloor$. $d_{asym}$ can also be found with similar averaging using $\textrm{AFD}^u_{m}$ instead of $\textrm{FD}^u_{m}$.
%\begin{eqnarray}
%\textrm{PSQM}_s^u=\sqrt[6]{\frac{1}{20}\sum_{i=0}^{19}\left( \textrm{FD}^u_{10s+i} \right)^6} \\
%d_{sym}=\sum_{u=0}^{B-1} \sqrt{\frac{1}{S^u}\sum_{s=0}^{S^u-1}\left( \textrm{PSQM}_s^u \right)^2}
%\end{eqnarray}
%where $S^u$ is $\lfloor \frac{M^u}{10} \rfloor$. $d_{asym}$ can also be found with similar averaging.

\begin{table*}[ht]
\caption{SDR and PESQ results on QUT-NOISE-TIMIT: Test set consists of 6 SNR ranges:-10, -5, 0, 5, 10, 15 dB. The highest SDR or PESQ scores for each SNR test data were highlighted with bold fonts.}
\label{tab:tq}
%\vskip 0.1in
\centering
%\begin{tabularx}{\textwidth}{l r r r r r r | r r r r r r }
\begin{tabular}{l r r r r r r | r r r r r r }
%\begin{tabular}{l r r r r r r | r r r r r r r}
\toprule
	& \multicolumn{6}{c|}{SDR}	& \multicolumn{6}{c}{PESQ} \\
%\midrule
Loss Type	& -10dB & -5dB & 0dB & 5dB & 10dB & 15dB & -10dB & -5dB & 0dB & 5dB & 10dB & 15dB\\
\midrule
Noisy Input & -11.82 & -7.33 & -3.27 & 0.21 & 2.55 & 5.03 & 1.07 & 1.08 & 1.13 & 1.26 & 1.44 & 1.72 \\
IAM & -3.23  & 0.49 & 2.79 & 4.63 & 5.74 & 7.52 & 1.29 & 1.47 & 1.66 & 1.88 & 2.07 & 2.30 \\ 
PSM & -2.95 & 0.92 & 3.37 & 5.40 &6.64 & 8.50 & 1.30 & 1.49 & 1.71 & 1.94 & 2.15 & 2.37 \\
%SNR & -2.79 & 1.36 & 3.70 & 5.68 & 6.18 & 8.44 & 1.29 & 1.46 & 1.68 & 1.93 & 2.14 & 2.38 \\
SDR & -2.66 & 1.55 & 4.13 & 6.25 & 7.53 & 9.39 & 1.26 & 1.42 & 1.65 & 1.92 & 2.16 & 2.41  \\
%SDR-MSE & -2.53 & 1.57 & 4.10 & 6.31 & 7.60 & 9.43 & 1.29 & 1.47 & 1.68 & 1.93 & 2.15 & 2.39 \\
%SI-SDR-PESQ* & -2.20 dB & 1.82 dB & 4.27 dB & 6.30 dB & 7.43 dB & 9.20 dB & 1.46 & 1.67 & 1.90 & 2.16 & 2.38 & 2.62 \\
SDR-STOI & -2.45 & 1.67 & 4.22 & 6.38 & 7.62 & 9.56 & 1.30 & 1.49 & 1.72 & 1.99 & 2.21 & 2.49 \\
SDR-PESQ & \textbf{-2.31} & \textbf{1.80} & \textbf{4.36} & \textbf{6.51} & \textbf{7.79} & \textbf{9.65} & \textbf{1.43} & \textbf{1.65} & \textbf{1.89} & \textbf{2.16} & 2.35 & 2.54   \\  
%SDR-PESQ-STOI & -2.23 & 1.85 & 4.36 & 6.48 & 7.72 & 9.59 & 1.48 & 1.71 & 1.96 & 2.25 & 2.47 & 2.71 \\
SDR-PESQ-STOI & -2.34 & 1.78 & 4.22 & 6.37 & 7.54 & 9.36 & 1.37 & 1.59 & 1.83 & 2.13 & \textbf{2.37} & \textbf{2.61} \\
\bottomrule
\end{tabular}
\end{table*}

\setlength{\tabcolsep}{3pt}
\begin{table}[ht]
\caption{STOI results on QUT-NOISE-TIMIT}
\label{tab:stoi}
%\vskip 0.1in
\centering
%\begin{tabularx}{\textwidth}{l r r r r r r | r r r r r r }
\begin{tabular}{l r r r r r r}
\toprule
%	& \multicolumn{5}{c}{STOI}	 \\
%\midrule
Loss Type	& -10dB & -5dB & 0dB & 5dB & 10dB & 15dB \\
\midrule
Noisy Input & 51.9 & 60.4 & 68.6 & 76.9 & 82.4 & 86.9\\
SDR & 58.5 & 70.8 & 79.0 & 84.6 & 87.3 & 89.6 \\
SDR-PESQ & 60.6 & 72.6 & 80.4 & 85.6 & 88.2 & \textbf{90.3}\\
SDR-STOI & 60.6 & 72.5 & 80.2 & 85.5 & 88.1 & \textbf{90.3}\\
SDR-PESQ-STOI & \textbf{60.8} & \textbf{73.0} & \textbf{80.6} & \textbf{85.7} & \textbf{88.3} & \textbf{90.3}\\
\bottomrule
\end{tabular}
\end{table}

\begin{table}[ht]
\caption{Evaluation on VoiceBank-DEMAND corpus}
\label{tab:ge}
%\vskip -0.1in
\centering
\begin{tabularx}{\columnwidth}{l X X X X X X}
%\begin{tabular}{l r r r r r r}
\toprule
Models	& CSIG	& CBAK	& COVL	& PESQ	& SSNR & SDR\\
\midrule
Noisy Input	&	3.37	&	2.49	&	2.66	&	1.99	&	2.17 & 8.68\\
SEGAN		&	3.48	&	2.94	&	2.80	&	2.16	&	7.73 & -\\
WAVENET		&	3.62	&	3.23	&	2.98	&	-		&	-	& - \\
TF-GAN		&	3.80	&	3.12	&	3.14	&	2.53	&	- & - \\
SDR-PESQ (ours)	&	\textbf{4.09}	&	\textbf{3.54}	&	\textbf{3.55}	&	\textbf{3.01}	&	\textbf{10.44} & \textbf{19.14}\\
%TF G.W.	&	\textbf{4.18}	&	\textbf{3.59}	&	\textbf{3.62}	&	\textbf{3.06}	&	\textbf{10.78} & \textbf{19.57} \\
\bottomrule
\end{tabularx}
\end{table}

\subsection{STOI Loss Function} %ME: same commment about "`the"'
\begin{figure}[ht]
%\begin{figure*}
\vskip -0.2in
\begin{center}
\centerline{\includegraphics[width=70mm]{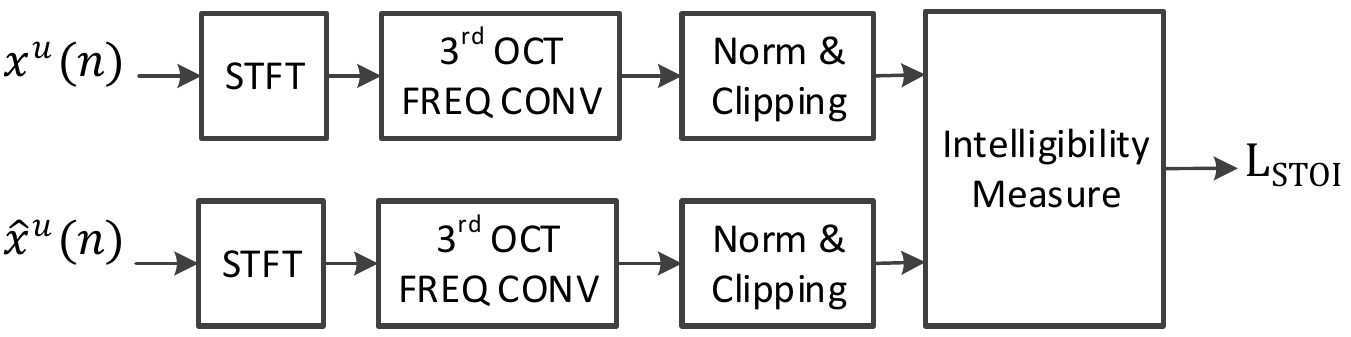}}
%\centerline{\includegraphics[width=\textwidth]{figs/denoising_framework-cropped.pdf}}
\caption{Block Diagram for the $\textrm{STOI}$ loss function}
\label{fig:stoi}
\label{fig:stoi}
\end{center}
\vskip -0.4in
\end{figure}
STOI~\cite{taal2011algorithm, zhao2018perceptually} is a segmented correlation-based metric. STOI optimized  the segment length that provides the highest correlation with the speech intelligibility. The experiment result of Figure~10 in~\cite{taal2011algorithm} showed STOI gets the highest correlation for segment lengths around hundreds of milliseconds. The analysis window length of $384$ ms is chosen based on this criterion.

The proposed STOI loss function is depicted in Figure~\ref{fig:stoi}. The clean and denoised signal pairs are converted into complex spectra vis STFT. After linear frequency scale is converted into 1/3-octave frequency scale, a noisy spectrum is normalized and clipped by the norm of a clean spectrum. The intelligibility measure is calculated by correlation between normalized clean and noisy spectra. Finally, $\textrm{L}_{\textrm{STOI}}$ is averaged over 1/3 octave bands, frames and mini-batch utterances:
\begin{equation}
\textrm{L}_{\textrm{STOI}} = \frac{1}{BMJ}\sum_{u=0}^{B-1}\sum_{m=0}^{M-1}\sum_{j=0}^{J-1}d_{m,j}^u
\end{equation} 
where B is the number of mini-batch utterances, M is the number of frames, J is the number of 1/3 octave bands and $d_{m,j}^u$ is intelligibility measure at utterance $u$, frame $m$ and 1/3-octave band $j$ which is described in detail at Eq. (5) in~\cite{taal2011algorithm}.

\subsection{Joint optimization of perceptual speech metrics} %ME: iS it better to call it Joint SDR, PESQ, and STOI optimization
\label{sec:joint}
In this section, several new loss functions are defined by combining SDR, PESQ and STOI loss functions. The first one is to jointly maximize SDR and PESQ by
combining $L_{\textrm{SDR}}$ and $\textrm{L}_{\textrm{PESQ}}$:
\begin{equation}
\textrm{L}_{\textrm{SDR-PESQ}}=\textrm{L}_{\textrm{SDR}}+\alpha \textrm{L}_{\textrm{PESQ}}
\end{equation}
where $\alpha$ is a hyper-parameter to adjust relative weighting between SDR and PESQ loss functions.

Similarly, SDR and STOI can be jointly maximized by combining $L_{\textrm{SDR}}$ and $\textrm{L}_{\textrm{STOI}}$ as follows:
\begin{equation}
\textrm{L}_{\textrm{SDR-STOI}}=\textrm{L}_{\textrm{SDR}}+\beta \textrm{L}_{\textrm{STOI}}
\end{equation}
where $\beta$ is a hyper-parameter to adjust relative weighting between SDR and STOI loss functions.

%Another combined loss function can be defined using MSE criterion instead of $\textrm{L}_{\textrm{PESQ}}$ or $\textrm{L}_{\textrm{STOI}}$:
%\begin{eqnarray}
%&&\textrm{L}_{\textrm{SDR-MSE}}=\textrm{L}_{\textrm{SDR}}+ \nonumber \\
%&&\alpha \sum_{u=0}^{B-1}\sum_{m=0}^{M-1}\sum_{k=0}^{K-1} \left( M^u_{m,k}|Y^u_{m,k}| -|X^u_{m,k}| \right)^2
%\end{eqnarray}
%By comparing $\textrm{L}_{\textrm{SDR-MSE}}$ with $\textrm{L}_{\textrm{SDR-PESQ}}$ or $\textrm{L}_{\textrm{SDR-STOI}}$, we can evaluate PESQ or STOI improvement from the proposed PESQ or STOI loss functions over MSE criterion. 

%We also define two other multi-task loss functions for the joint optimization of the SDR and STOI, as well as the joint optimization of SDR, PESQ, and STOI together, as defined below:
Finally, we can combine all three loss functions to jointly optimize SDR, PESQ and STOI metrics as follows:
\begin{equation}
\textrm{L}_{\textrm{SDR-PESQ-STOI}}=\textrm{L}_{\textrm{SDR}}+\alpha \textrm{L}_{\textrm{STOI}}+\beta \textrm{L}_{\textrm{STOI}}
\end{equation}

\section{Experimental Results}
\label{sec:exp}

\subsection{Experimental Settings}
Two datasets were used for the evaluation of the proposed denoising framework. QUT-NOISE-TIMIT~\cite{dean2010qut} is synthesized by mixing 5 different background noise sources with the TIMIT~\cite{garofolo1993darpa}. For the training set, -5 and 5 dB SNR data were used but the evaluation set contains all SNR ranges. The total length of train and test data corresponds to 25 hours and 12 hours, respectively. For VoiceBank-DEMAND~\cite{valentini2016investigating}, 30 speakers selected from Voice Bank corpus~\cite{veaux2013voice} were mixed with 10 noise types:8 from Demand dataset~\cite{thiemann2013diverse} and 2 artificially generated one. Test set is generated with 5 noise types from Demand.

\subsection{Main Result}
\label{sec:main}

Table~\ref{tab:tq} compared SDR and PESQ performance between different denoising methods on QUT-NOISE-TIMIT corpus. All the schemes were based on the same CNN-BLSTM model trained with -5 and +5 dB SNR data. IAM and PSM are the existing spectral mask estimation schemes. SDR refers to $\textrm{L}_{\textrm{SDR}}$ at Section~\ref{sec:si-sdr} and SDR-PESQ, SDR-STOI and SDR-PESQ-STOI correspond to $\textrm{L}_{\textrm{SDR-PESQ}}$, $\textrm{L}_{\textrm{SDR-STOI}}$ and $\textrm{L}_{\textrm{SDR-PESQ-STOI}}$ at Section~\ref{sec:joint}, respectively. 

$\textrm{L}_{\textrm{SDR}}$ outperformed IAM and PSM for all SNR ranges in terms of SDR metric. However, for PESQ, it did not show similar improvement due to the metric mismatch. $\textrm{L}_{\textrm{SDR-PESQ}}$ loss function improved both SDR and PESQ metrics. $\textrm{L}_{\textrm{PESQ}}$ loss function acted as a regularization term to improve not only PESQ but also the SDR metric. The STOI loss function was not as effective as PESQ loss function to improve PESQ and SDR metrics. SDR-STOI and SDR-PESQ-STOI showed degradation on the SDR metric compared with SDR-PESQ.   

Table~\ref{tab:stoi} compared STOI performance among different loss functions. $\textrm{L}_{\textrm{SDR-STOI}}$ loss function showed
$1.52$\% relative STOI gain over the SDR loss function. One thing to note is that $\textrm{L}_{\textrm{SDR-PESQ}}$ showed the similar improvement on STOI as $\textrm{L}_{\textrm{SDR-SI}}$, which is a surprising result because the model based on $\textrm{L}_{\textrm{SDR-PESQ}}$ was not directly trained with STOI function but its generalization on the STOI metric is as good as the model based on $\textrm{L}_{\textrm{SDR-STOI}}$. We further evaluated STOI performance by combining all three loss functions. However, $\textrm{L}_{\textrm{SDR-PESQ-STOI}}$ showed only marginal STOI improvements over $\textrm{L}_{\textrm{SDR-PESQ}}$. Considering the result on SDR and PESQ metrics, $\textrm{L}_{\textrm{SDR-PESQ}}$ loss function showed the most consistent performance over all three metrics. 
%\subsection{Evaluation on Complex Transformer Model}
\subsection{Comparison with Generative Models}
%\vspace{-2em}

Table~\ref{tab:ge} showed comparison with other generative models. All the results except our end-to-end model came from the original papers: SEGAN~\cite{pascual2017segan}, WAVENET~\cite{rethage2018wavenet} and TF-GAN~\cite{soni2018time}. CSIG, CBAK and COVL are objective measures, where high value means better quality of speech~\cite{hu2008evaluation}. CSIG is mean opinion score (MOS) of signal distortion, CBAK is MOS of background noise intrusiveness and COVL is MOS of the overall effect. SSNR is Segmental SNR defined in~\cite{quackenbush1986objective}.

The proposed SDR and PESQ joint optimization scheme outperformed all the generative models in all the perceptual speech metrics listed above. One thing to note is that any of the metrics at Table~\ref{tab:ge} was not used as a loss function but the proposed SDR and PESQ combined loss function is highly effective to those metrics, which suggested its good generalization performance.
%Table~\ref{tab:ge} showed comparison with other generative models. All the results except CNN-LSTM and TF G.W. came from the original papers: SEGAN~\cite{pascual2017segan}, WAVENET~\cite{rethage2018wavenet} and TF-GAN~\cite{soni2018time}. CSIG, CBAK and COVL are objective measures where high value means better quality of speech~\cite{hu2008evaluation}. CSIG is mean opinion score (MOS) of signal distortion, CBAK is MOS of background noise intrusiveness and COVL is MOS of the overall effect. SSNR is Segmental SNR defined in~\cite{quackenbush1986objective}. 
%
%The proposed Transformer model outperformed all the generative models for all the perceptual speech metrics listed in Table~\ref{tab:ge} with large margin. The main improvement came from the joint SDR and PESQ optimization schemes in~\cite{kim2019end} that benefited both CNN-LSTM and TF G.W. However, TF G.W showed consistently better performance over CNN-LSTM for all the metrics, which agreed with the result at Table~\ref{tab:tq}.
\section{Conclusion}
\label{sec:related}
In this paper, a new end-to-end multi-task denoising scheme was proposed. The proposed scheme resolved two issues addressed before: spectrum and metric mismatches.
%First, three metric-based loss functions are defined: SDR, PESQ and STOI loss functions. Second, three newly defined loss functions are combined for the joint SDR, PESQ and STOI optimization. Finally, the combined loss function is optimized based on the reconstructed time-domain signal after Griffin-Lim ISTFT in order to avoid spectrum mismatch. 
The experimental result presented that the proposed joint optimization scheme significantly improved SDR, PESQ and STOI performances over both spectral mask estimation schemes and generative models. Moreover, the proposed scheme provided good generalization performance by showing substantial improvement on the unseen perceptual speech metrics.

\bibliographystyle{IEEEtran}

\bibliography{mybib}

% \begin{thebibliography}{9}
% \bibitem[1]{Davis80-COP}
%   S.\ B.\ Davis and P.\ Mermelstein,
%   ``Comparison of parametric representation for monosyllabic word recognition in continuously spoken sentences,''
%   \textit{IEEE Transactions on Acoustics, Speech and Signal Processing}, vol.~28, no.~4, pp.~357--366, 1980.
% \bibitem[2]{Rabiner89-ATO}
%   L.\ R.\ Rabiner,
%   ``A tutorial on hidden Markov models and selected applications in speech recognition,''
%   \textit{Proceedings of the IEEE}, vol.~77, no.~2, pp.~257-286, 1989.
% \bibitem[3]{Hastie09-TEO}
%   T.\ Hastie, R.\ Tibshirani, and J.\ Friedman,
%   \textit{The Elements of Statistical Learning -- Data Mining, Inference, and Prediction}.
%   New York: Springer, 2009.
% \bibitem[4]{YourName17-XXX}
%   F.\ Lastname1, F.\ Lastname2, and F.\ Lastname3,
%   ``Title of your INTERSPEECH 2020 publication,''
%   in \textit{Interspeech 2020 -- 20\textsuperscript{th} Annual Conference of the International Speech Communication Association, September 15-19, Graz, Austria, Proceedings, Proceedings}, 2020, pp.~100--104.
% \end{thebibliography}

\end{document}